\documentclass[aps,pre,twocolumn,groupedaddress]{revtex4}

\usepackage{amsmath}    % need for subequations
\usepackage{graphicx}   % need for figures
\usepackage{verbatim}   % useful for program listings
\usepackage{color}      % use if color is used in text
\usepackage{subfigure}  % use for side-by-side figures
\usepackage{rotating}
\usepackage{footnote}

\begin{document}

\title{The Timing of Sonoluminescence}
\author{Thomas E. Brennan}
\affiliation{}
\author{Gustave C. Fralick}
\affiliation{NASA Glenn Research Center}
\date{\today}

\begin{abstract}

We measured the timing of sonoluminescence by observing laser light scattered from a single sonoluminescing bubble.  We performed this measurement on 23.5 kHz, 17.8 kHz, 13.28 kHz and 7920 Hz systems, and found that the flash typically occurs 100 nanoseconds before the minimum radius.  These results motivate a new model of sonoluminescence:  the flash results from the discharge of an excited cold condensate formed during the adiabatic expansion of the bubble.

\end{abstract}

\maketitle

\section*{Introduction}

Sonoluminescence is the transformation of sound into light.  In practice, sonoluminescence is created by focusing powerful sound waves in water or other fluids.  Small bubbles present in the fluid will glow if the sound field is strong enough, and other conditions are appropriate: such as the amount and type of dissolved gas present and the ambient temperature of the fluid.  The glow appears continuous to the eye, but in fact is the result of a rapid flashing which occurs once per acoustic cycle.

A determination of the timing of the sonoluminescence flash in the acoustic cycle is essential
to forming a causal picture of the phenomenon.  That is, to answer the question {\bf ``Why does the bubble glow?"}  

When experimenters first developed the technique to trap a single bubble in an acoustic field, \cite{GaitanThesis}, it became possible to study the time evolution of the expansion and contraction cycle of the bubble wall, and the timing of the sonoluminescence flash with respect to that cycle. By observing laser light scattered from a glowing bubble with a photomultiplier tube and oscilloscope, the expansion and contraction of the bubble as a function of time can be recorded with an oscilloscope; the sonoluminescence flash then appears as a narrow spike superimposed on the bouncing bubble pattern.

What is found is that during the cycle the bubble expands, reaches a maximum radius,
then rapidly collapses to a minimum and bounces several times.  The cycle then repeats.
The flash occurs only once per cycle, just before the time of the first minimum radius.  Exactly how close in time to this first minimum is the object of this experimental investigation.

A series of studies of the timing of the flash in the bubble cycle were published beginning
in 1991, \cite{BarberAndPutterman}, \cite{PuttermanMieScattering}, \cite{PuttermanTimeCorrelated}, by one group which claimed that the sonoluminescence flash always occurs within a fraction of a nanosecond of the minimum bubble radius. These studies claimed to have sub-nanosecond resolution. However, only a handful of actual oscilloscope traces were published by this group showing the sonoluminescence flash and the bubble wall motion on the same trace. These claims were backed up by others, \cite{MatulaAndCrum}, \cite{CrumEditor}, \cite{GompfAndEisenmenger}, but those articles do not show oscilloscope traces zoomed-in near the minimum radius.

These studies that claim to show the flash occurring synchronously with the minimum bubble radius support {\bf the ÔhotÕ paradigm of sonoluminescence:
that the violent collapse of the bubble wall is the ultimate cause of the sonoluminescence
flash.} In this model, the supersonic collapse of the bubble wall launches a spherically imploding shockwave that heats the gases in the bubble into a short-lived plasma flash. According
to a numerical model put forward by Greenspan, \cite{Greenspan}, and Wu, \cite{WuSimulation}, the extreme conditions leading to the flash set in less than one tenth of one nanosecond before the minimum bubble radius.  In other words, the hot theory of sonoluminescence requires the flash to occur nearly synchronously with the time of the first minimum radius.

Also, studies of the spectrum of the emitted light can be interpreted as supporting
this hot paradigm.  Sonoluminescence produces a continuous spectrum resembling a black
body, and numerous studies, \cite{HillerThesis}, \cite{GaitanCalibrated}, show that fits to black body or bremsstrahlung curves imply temperatures in the bubble at the moment of the flash exceeding $10^4$ K, perhaps as high as $10^6$ K.

Thus it would seem that the case regarding the cause of sonoluminescence is closed:
published measurements of the timing and spectrum of sonoluminescence confirm the details
of the hot/shockwave paradigm. But for one objection: only one group has ever published oscilloscope traces of the flash which are zoomed-in at the nanosecond scale near the minimum radius, and those that have been published are not clear and are not numerous.

We decided to repeat the measurements of the timing of the sonoluminescence flash.

\section*{Experimental Design and Procedure}

\begin{figure}[htp!]
\centering
\includegraphics[scale=0.6]{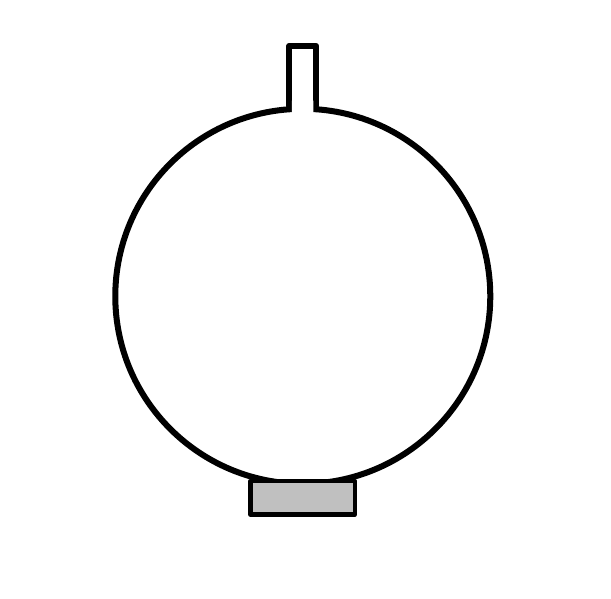}
\caption{\label{fig:SLResonator}  The design of the 500 mL, 1 L and 5 L resonators.}
\end{figure}

Each of the three resonator flasks we constructed is a hollow quartz sphere with a narrow neck and thin walls, (see Figure \ref{fig:SLResonator}).  The 500 mL flask was fitted with a one inch cylindrical PZT ring epoxied to the bottom center, and was held  on the test stand by the neck with a small lab clamp.  The 1 L flask was fitted with a two inch cylindrical PZT ring.  Because of its larger weight, it was not suspended by a lab clamp, but by a rubber hose epoxied to the outside of the neck.  The 5 L flask was fitted with a three inch cylindrical PZT and suspended by a rubber hose as well.  All three flasks also had small PZT disks epoxied to the side wall which functioned as microphones to monitor the response of the flask to the drive.

The experiment begins by preparing the water-glycerine-gas mixture.  It is important to minimize the amount of dust in the fluid mixture, because dust particles interfere with the formation of the  bubble. So in a clean room, we mix approximately 1.36 liters of glycerine and 6.63 liters of deionized water for a total fluid volume of 8 liters, in a large 9.5 liter mixing flask.  We cap the flask with the rubber stopper, and then take it out of the clean room and bring it to the lab room where we do the rest of the experiment.  (This lab room could be made completely dark, and was cooled to 15 Celcius.)  We then degas the mixture for several hours with a vacuum pump while stirring.  After the fluid has been degassed, we add about 1/5th atm of argon gas above the fluid, and allow it to stir vigorously for at least 24 hours.  After this, the fluid has been infused with argon and is ready to be used.

Next, the fluid is poured into the resonator.  To do this, the top of the mixing flask is opened to allow the fluid to flow thru a stopcock and tube into the flask.  After the pouring is complete, the vacuum pump is used to remove the air above the fluid, the low pressure argon atmosphere is restored, and the remaining fluid is stored for later use.

The resonator flask is then placed on the test stand, and the audio power and microphone leads are connected.  The first step is to find the exact resonant frequency of the flask so that bubbles can be levitated in the center of the flask.  From a priori calculations, we know approximately what this frequency should be, so we set the generator to this frequency, set the drive of the generator to about 1.5 Vpp, and then ramp up the power amplifier to about one third of max.  Next, we adjust the value of the inductance to optimize the impedance match between the amplifier and the resonant circuit.  Then we turn the power amplifier up to about two-thirds of max, and attempt to trap a bubble by disturbing the surface of the water with a syringe.  We continue to adjust the drive and frequency until a bubble can be made to levitate in the center of the flask and cavitate.  (When a bubble begins to cavitate it gets a ``fuzzy" appearance).   At this point, we turn the room lights off and make further minute adjustments to the frequency and drive until the bubble begins to glow visibly.  The 500 mL resonator has a resonant frequency close to 17.8 kHz, the 1 L resonator has a resonant frequency close to 13.28 kHz, and the 5 L resonator has a resonant frequency close to 7920 Hz. 

\begin{figure}[htp!]
\centering
\includegraphics[scale=0.6]{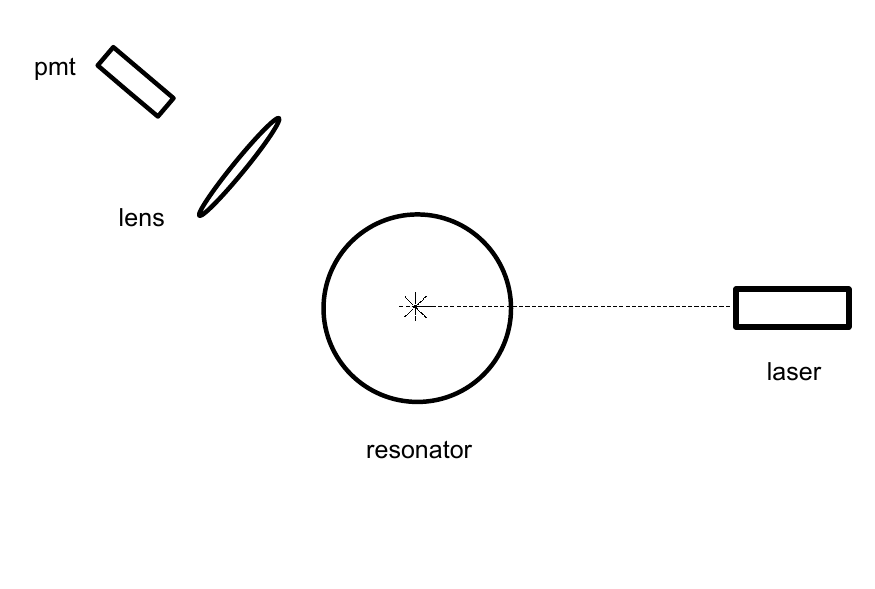}
\caption{\label{fig:SLScatteringSchematic}  A schematic of the laser scattering experiment.}
\end{figure}

To study the oscillation of the bubble and the timing of the flash in the bubble cycle, we scattered laser light from the bubble and monitored the scattered light signal with a lens, photomultiplier tube and oscilloscope (see Figure \ref{fig:SLScatteringSchematic}).  This experiment is similar to previous Mie-scattering experiments, \cite{PuttermanMieScattering}, \cite{myThesis}, \cite{mySLArticle}.  The laser we used was a 0.95 mW red HeNe.  A large 4.5 inch diameter objective lens with a 10 cm focal length  was used to focus scattered light from the bubble onto a Hamamatsu R647-25 PMT with a 2.5 ns rise time.  The PMT was run at -850 Volts.  The microphone PZT, sync signal from the generator, and the output of the PMT were all monitored and recorded with a Tektronix TDS 5034B oscilloscope which has a sampling rate as high as 3 GHz.

\section*{Results}

Based upon repeated observations we find that the sonoluminescence flash typically occurs 100 nanoseconds before the time of minimum bubble radius.  The sequence of events is as follows:  the bubble at first maintains a steady radius for about 50\% of the period.  Then it suddenly expands to a larger radius, reaches an apex, then begins to collapse.  While collapsing, it emits a brief flash of light; after that light is extinguished, the bubble continues to shrink for an additional 100 nanoseconds, and sometimes longer.  Finally, the bubble reaches a distinct minimum radius, rebounds, bounces several times, and then stabilizes.  The cycle repeats from there.  

See Figures \ref{fig:110706_001063} thru \ref{fig:09112006dZ}  for examples.  We have left these oscilloscope traces as pure data, unaugmented with any markings except the scale of the volage and time axis.  We encourage the reader to take a straight edge, obtain the scale, and verify our claims.  Note that the photomultiplier tube that we used outputs a negative voltage which is proportional to the cross-sectional area of the bubble, so that these graphs are inverted.

\section*{17.8 kHz Data}

\begin{turnpage}
\begin{figure}[htp!]
\centering
\includegraphics[scale=1.0, angle=0]{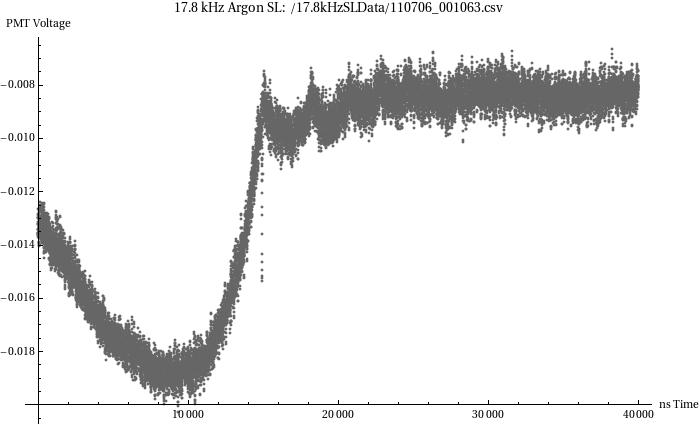}
\caption{\label{fig:110706_001063} One cycle of a 17.8 kHz argon SL.  }
\end{figure}
\end{turnpage}

\begin{turnpage}
\begin{figure}[htp!]
\centering
\includegraphics[scale=1.0, angle=0]{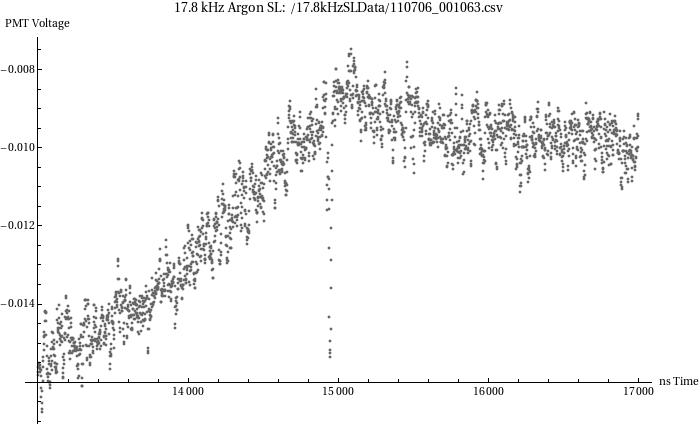}
\caption{\label{fig:110706_001063Z}  Zoom-in on previous. }
\end{figure}
\end{turnpage}

Figure \ref{fig:110706_001063} depicts one cycle of 17.28 kHz sonoluminescence.  It is an average of eight traces.  In this figure, several bounces are clearly visible, of at least three distinct minimum radii, showing that the motion of the bubble is stable over several periods and that timing information is not lost due to oscilloscope averaging.  Thus when we look at Figure \ref{fig:110706_001063Z}, which is a zoom-in near the first bounce, we see that the sonoluminescent flash, visible as the upside down spike, precedes the minimum radius by over 100 nanoseconds.  Since this spike is only slightly broader than the 10 nanosecond width of a single-shot trace of a flash on our system, \cite{mySLArticle}, we know that we see compelling evidence of several flashes that repeatedly and regularly occurred over 100 nanoseconds before the minimum radius.

\begin{turnpage}
\begin{figure}[htp!]
\centering
\includegraphics[scale=1.0, angle=0]{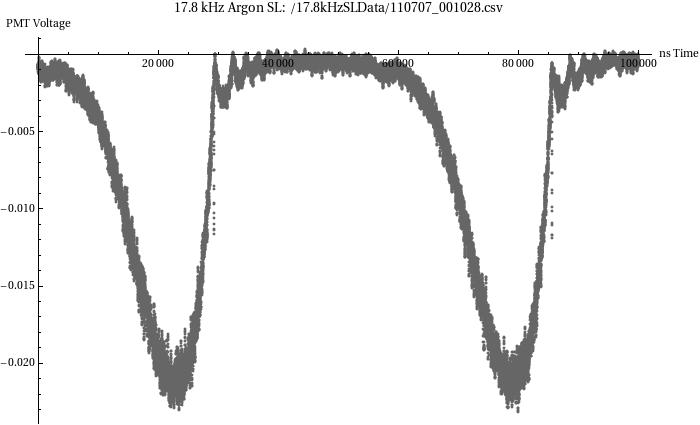}
\caption{\label{fig:110707_001028}  Two cycles of a 17.8 kHz argon SL. }
\end{figure}
\end{turnpage}

\begin{turnpage}
\begin{figure}[htp!]
\centering
\includegraphics[scale=1.0, angle=0]{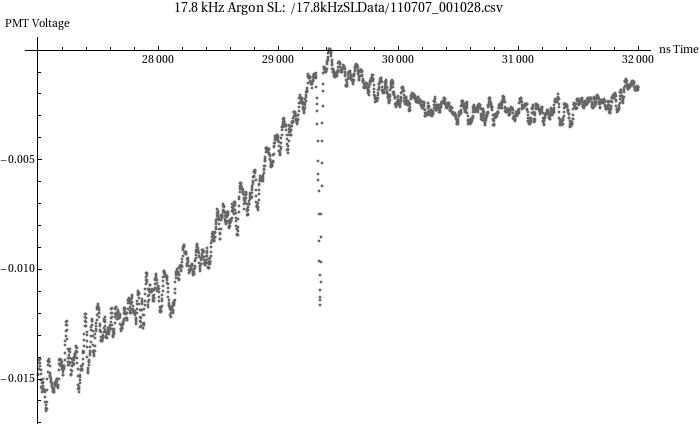}
\caption{\label{fig:110707_001028Z}  Zoom-in on first flash in previous. }
\end{figure}
\end{turnpage}

Figure \ref{fig:110707_001028} shows two full bubble periods.  The zoom-in in Figure \ref{fig:110707_001028Z} is on the first flash, which is somewhat broadened due to averaging, but still distinct from and preceding the minimum radius by at least 100 nanoseconds.

\section*{13.28 kHz Data}

\begin{turnpage}
\begin{figure}[htp!]
\centering
\includegraphics[scale=1.0, angle=0]{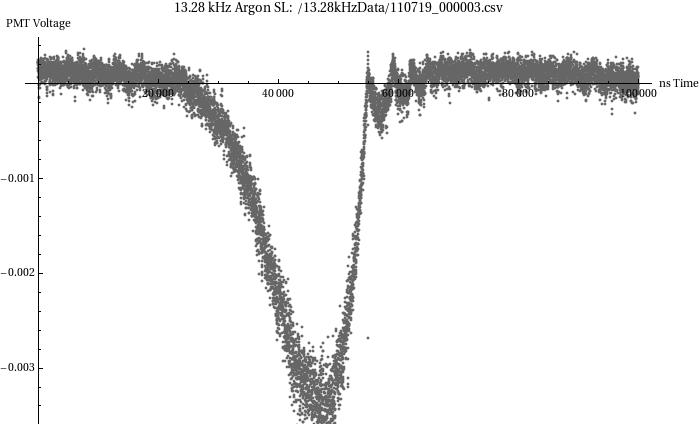}
\caption{\label{fig:110719_000003}  One cycle of 13.28 kHz argon SL. }
\end{figure}
\end{turnpage}

\begin{turnpage}
\begin{figure}[htp!]
\centering
\includegraphics[scale=1.0, angle=0]{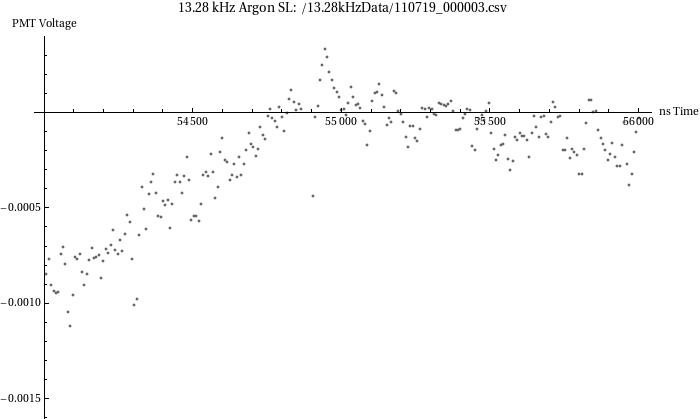}
\caption{\label{fig:110719_000003Z}  Zoom-in on previous. }
\end{figure}
\end{turnpage}

Figure \ref{fig:110719_000003}  shows the advantage of viewing the data on a large scale before zooming in near the minimum radius.  Notice how well defined the first minimum radius bounce is.  Three or four data points appear to point to it distinctly.  We should keep this in mind when we look at the zoom-in shown in Figure \ref{fig:110719_000003Z}, which at first appears less definitive, unless we identify the three points of the minimum radius from the zoom out, measure the difference between the center of these three points and the beginning of a broad flash which exceeds the scale of the zoom.  Again, a time difference in excess of 100 nanoseconds is apparent.

\section*{7920 Hz Data}

\begin{turnpage}
\begin{figure}[htp!]
\centering
\includegraphics[scale=1.0, angle=0]{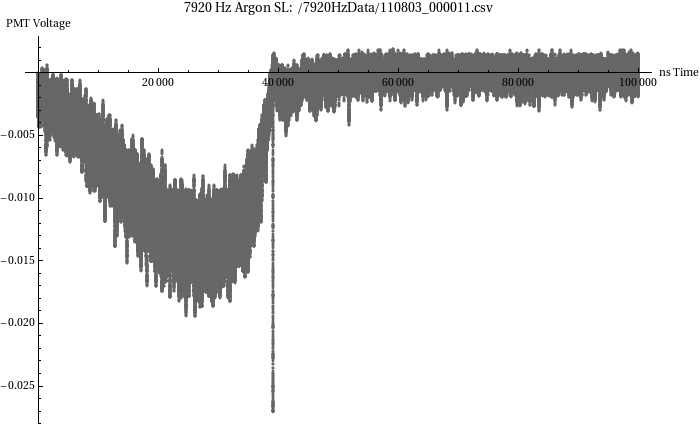}
\caption{\label{fig:110803_000011}  One cycle of 7920 Hz argon SL. }
\end{figure}
\end{turnpage}

\begin{turnpage}
\begin{figure}[htp!]
\centering
\includegraphics[scale=1.0, angle=0]{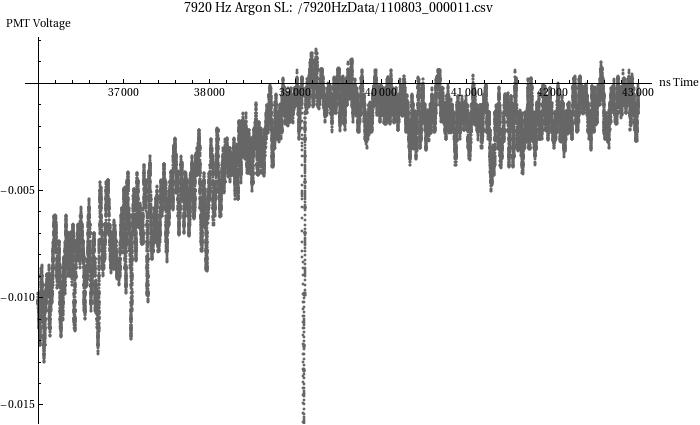}
\caption{\label{fig:110803_000011Z}  Zoom-in on previous. }
\end{figure}
\end{turnpage}

This trace in Figure \ref{fig:110803_000011} depicts data from a single shot oscilloscope trace.  Looking at the full image, we can see that it appears noisy, but a zoom-in near the first bounce in Figure \ref{fig:110803_000011Z} reveals that the flash occurs about 100 nanoseconds before the minimum radius.

\section*{23.5 kHz Data}

\begin{turnpage}
\begin{figure}[htp!]
\centering
\includegraphics[scale=1.0, angle=0]{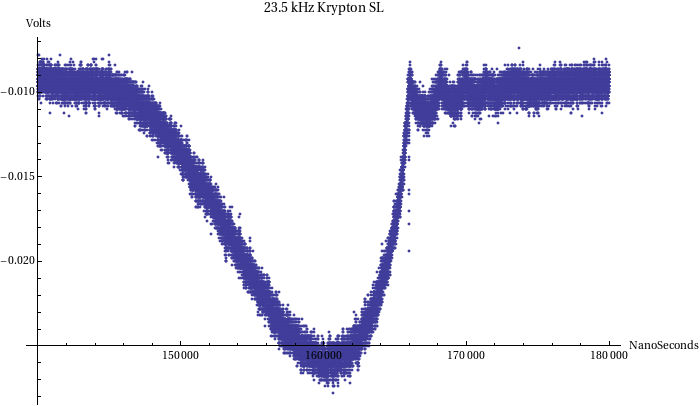}
\caption{\label{fig:09112006d}  One cycle of 23.5 kHz Krypton SL.  This is a single-shot oscilloscope trace }
\end{figure}
\end{turnpage}

\begin{turnpage}
\begin{figure}[htp!]
\centering
\includegraphics[scale=1.0, angle=0]{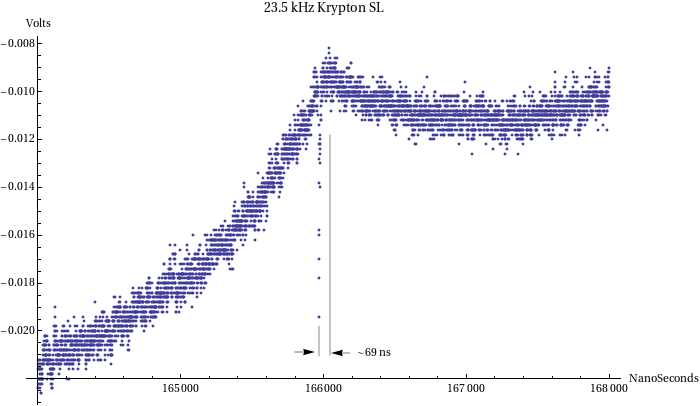}
\caption{\label{fig:09112006dZ}  A zoom-in near the minimum radius of the previous trace.  Note the distinct flash and minimum radius and the distinct time separation between them. }
\end{figure}
\end{turnpage}

Here we include a few oscilloscope traces of a 23.5 kHz sonoluminescence system which was constructed and tested as part of the thesis work of the first author of this article.  This data was taken in 2009 at the Illinois Institute of Technology using a very similar procedure to that performed in 2011 at NASA/Glenn.  We have chosen to include these traces, Figures \ref{fig:09112006d} and \ref{fig:09112006dZ}, of a sonoluminescing krypton bubble because they represent some of our cleanest data showing a distinct time separation between the flash and minimum radius.

\subsection*{Hypothesis of Condensate Energy}

These experimental results imply that the compressive-shock heating model of \emph{Greenspan}, \cite{Greenspan}, \emph{Wu}, \cite{WuSimulation}, or others like it that require the flash to occur within one nanosecond of the time of minimum bubble radius must be ruled out.  We present a model which is consistent with the fact that the flash occurs 100 nanoseconds before the minimum bubble radius:  a time when the bubble wall motion is subsonic, and temperatures and pressures are crossing through standard conditions.

\begin{figure}[htp!]
\centering
\includegraphics[scale=0.5]{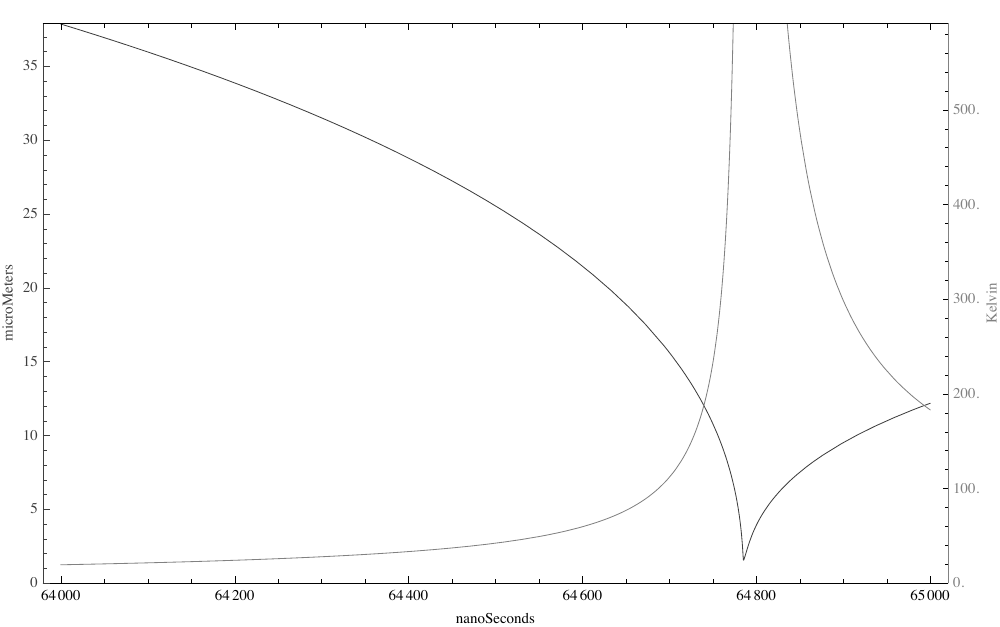}
\caption{\label{fig:XenonTempZoom}  A plot of bubble radius and temperature, based upon a solution of the Rayleigh-Plesset equation for a monatomic van der Waals gas.  The bubble temperature crosses thru 100 K approximately 100 nanoseconds before the time of minimum bubble radius, implying that any condensate formed during the expansion will be evaporated by this time.}
\end{figure}

Let's consider the temperature fluctuations which occur in the bubble as a result of its expansion and contraction.  The calibration of the laser scattering data as reported in previous experiments, \cite{PuttermanReview}, \cite{BrennerReview}, \cite{PuttermanMieScattering}, \cite{mySLArticle}, \cite{myThesis},  implies that the maximum radius of the bubble is larger by a factor of ten compared to the equilibrium radius.  This is a volume expansion ratio of 1000.  Consider that during adiabatic expansion, the temperature of a van der Waals gas drops as $ 1/V^{(\gamma - 1)} $.   Assuming $\gamma$ to be $5/3$ such as for a monatomic gas like argon, then $T \propto V^{-2/3}$.  Thus, a radial expansion factor of ten will result in a hundred-fold decrease in temperature.  Therefore if the gas is at room temperature at the equilibrium radius, it will be cooled to a temperature below 4 Kelvin by the time it reaches the maximum radius -- a temperature cold enough to liquify or freeze any gases in the bubble.  When the bubble subsequently collapses, the contents of the bubble will warm back up, and any condensate formed during the expansion will be destroyed by the time the bubble returns to its equilibrium radius, which will occur approximately 100 ns before the time of minimum bubble radius, (see Figure \ref{fig:XenonTempZoom}).

One note at this point on a common objection to the assumption that thermodynamic equilibrium 
could prevail in the interior of the bubble on the acoustic time scales involved -- that is, prevail to the extent that adiabatic cooling and heating leading to phase transformations in the bubble could take place.  Our response to this objection is that the assumption of adiabaticity has precedent in acoustic theory:  the well known derivation of the speed of sound requires that on acoustic time scales, a differential parcel of gas or fluid is isolated from its contiguous surroundings and undergoes adiabatic expansion and compression as a sound wave passes.  Assuming adiabaticity results in the correct value for the speed of sound, whereas assuming isothermal conditions gives the wrong result.  We expect adiabatic isolation to be even stronger for the case of a bubble than for a contiguous fluid.  Thus, the existence of thermodynamic equilibrium inside a bubble is an implication of a well established theory of sound which makes the same assumptions of internal equilibrium and external isolation on acoustic time scales. 

During adiabatic expansion, the gas in the bubble will cool as a result of performing work on its surroundings.  The pressure of the gas will drop as
\begin{equation}
P(V) = P_0 \frac{(V_0 -nb) ^{\gamma}}{(V-nb)^{\gamma}}
\label{simpleAdiabatic}
\end{equation}
With this we can calculate the work done by the gas during an adiabatic expansion.  We expect the integration of $P(V)$ to infinity to be convergent, and that the amount of work that can be performed by a gas on its surroundings, the \emph{adiabatic work capacity}, is finite:  
\begin{equation}
\label{workCapacity}
W_{\text{adiabatic}} = \int_{V_0}^{\infty} P(V) \, dV  = \frac{c_v}{R_g} P_0 (V_0 - nb) 
\end{equation}
Here, $R_g$ is the ideal gas constant, $c_v$ is the molar constant volume heat capacity, $n$ is the number of moles, and $b$ is the van der Waals excluded volume per mole.  The term  $ \frac{c_v}{R_g} P_0 V_0 $ is the work that could be performed by an ideal gas, which is reduced by the van der Waals term $ \frac{c_v}{R_g} P_0 n b $, the energy reserved for condensation. 

Let us discuss this concept of condensation energy further.  When a gas has been cooled to a temperature infinitesimally close to its condensation point, just before it condenses, there is energy present in the gas because the gas atoms are ``bouncing around" with center of mass motion.  Once it spontaneously condenses, it forms droplets which are relatively stationary about a single center of mass.  So then where did that energy go that was in the dispersed center of mass motion?  Well, it must be conserved, and to do so, the energy is transferred into excited electrons in the condensate.  The van der Waals empirical model implies that the term $ \frac{c_v}{R_g} P_0 n b $ represents the energy stored in this condensate.  For a bubble of equilibrium radius $R_0$ and temperature $ T_0 $ we can use the equation of state to replace the number of moles of gas, $ n $, in the condensate energy expression:
\begin{equation}
E_{\text{condensate}} = \frac{4 \pi}{3} \frac{c_v}{R_g} \frac{{P_0}^2 {R_0}^3}{R_g T_0} b
\label{usefulFormula}
\end{equation} 

Eq.\ (\ref{usefulFormula}) is a useful formula to the experimentalist.  It predicts that a $10 \mu m$ bubble starting at room temperature will form a condensate which stores approximately 1 picoJoule of energy.  Calibrated measurements of flashes from bubbles this size, \cite{GaitanCalibrated}, have shown that each flash contains about 1 picoJoule of energy.  Eq.\ (\ref{usefulFormula}) also predicts that cooler ambient temperatures will produce brighter bubbles, that larger bubbles will also be brighter, and that higher static pressure will produce brighter bubbles.  These predictions have been observed.

For a $10 \mu m$ bubble such as is typical, this formula predicts an energy of 1 picoJoule per flash, which is equivlent to 7 MeV.  On a macroscopic scale, 1 picoJoule doesn't seem like a lot of energy, but 7 MeV is a \emph{lot} of energy on the electron scale, especially if that 7 MeV is stored in one or a few electrons; in that case the discharge will certainly produce the bremsstrahlung or black-body spectra seen, \cite{HillerThesis}, as the discharge proceeds and thermalization sets in.

Note that this formula predicts that larger bubbles will have more energetic discharges.  If one were to trap $100 \mu m$ bubbles in larger, lower frequency resonators, condensate energy could exceed 7 GeV.  This might be enough energy to provoke a fusion reaction in the bubble if the right isotopes are present, in addition to producing a much more visibly powerful discharge.

To make the idea of condensation energy more tenable to the reader, we point out that it is essentially equivalent to the theory of \emph{latent heat}.  In order for a liquid fraction to evaporate, it must absorb a certain amount of heat: this is called the \emph{latent heat of vaporization}.  Likewise, when a portion of vapor condenses, it must shed this latent heat before rejoining the liquid phase.  Under normal conditions, the latent heat is shed invisibly in infrared modes of radiation.  But, under the special conditions of sonoluminescence, the focusing of latent heat occurs quickly enough to stimulate visible and ultraviolet excitations.

As a specific example,  consider a boiling kettle of water.  As the boiling proceeds, the energy of the blue flame underneath the kettle is transferred to the steam cloud above the kettle.  In principle, this light can be recovered if the steam is forced to re-condense quickly enough.  

We also point out that a form of low-frequency sonoluminescence might be behind the phenomenon of \emph{atmospheric lightning}.  The water vapor in summer thunderclouds contains an enormous amount of latent heat; if a large enough portion of a cloud were to suddenly condense into a meta-stable droplet, a great deal of energy would be focused and a large discharge could occur.  In other words, lightning \emph{is} sonoluminescence.  The low frequency \emph{boom} of thunder is the sound of the sudden collapse of a portion of a cloud into a condensate;  the higher frequency \emph{crack} of lightning is the sound produced by the high energy electrons discharging from the unstable condensate.  We note that lightning is known to emit broad-band electromagnetic radiation: from radio, through the visible and ultraviolet and into the $\gamma$ range, indicating the possibility of nuclear excitations.

\section*{Conclusion}

Based upon our understanding of thermodynamics, we believe that the rapid expansion of a sonoluminescing bubble suddenly cools its interior atmosphere to a few degrees Kelvin -- cool enough to condense the gases in the bubble.  Because of energy conservation, this sonically induced condensate must store the latent heat of the previous vapor in internal electronic excitations.  The condensate is meta-stable and persists until the collapsing bubble wall gets close enough for an electric arc discharge of several MeV to occur between the condensate and the bubble wall.    

We agree with the statement that sonoluminescence is an energy focusing phenomenon.  But our understanding is that the energy of the discharge is primed by the rapid cooling of the gas, rather than a compression shock;  for that mechanism requires high bubble wall velocities that are only expected to occur within one nanosecond of the minimum radius.  This is ruled out because we observe flashes at times 100 nanoseconds prior, when the bubble wall is still moving slowly.

One last comment about the absence of any statistical analysis of the flash timing in this work.  Our hypothesis aside, {\bf the \emph{thesis} of this article is a refutation of the claim that the sonoluminescence flash \emph{always} occurs within a nanosecond of the minimum radius.  To disprove that claim requires only a \emph{single} counterexample, while we have presented several.}  A statistical analysis of the mean and standard deviation would teach us more about how the timing varies, but it is not necessary for refuting the aforementioned claim.

%Data from November 2009

\end{document}